\documentclass[10pt]{article}
\usepackage{amsthm}
\usepackage{amssymb}
\usepackage{latexsym}
\newcommand{\Z}{{\mathbb Z}}

\newcommand{\R}{{\mathbb R}}

\def\be{\begin{equation}}
\def\ee{\end{equation}}
\def\bea{\begin{eqnarray}}
\def\eea{\end{eqnarray}}

\def\tr{{\rm \,tr\,}}

\def\q{{\bf q}}

\def\pv{{\bf p}}

\def\veps{\varepsilon}
\def\h2m{\frac{\hbar^2}{2m}}
\def\p0{{P_{\beta H^0_N}}}

\theoremstyle{nonumberplain}\newtheorem*{theorem}{Theorem}
\theoremstyle{nonumberplain}\newtheorem*{lemma}{Lemma}
\theoremstyle{nonumberplain}\newtheorem*{prop}{Proposition}

\begin{document}

\title{\large\bf Correlation inequalities for noninteracting Bose gases}
\author{Andr\'as S\"ut\H o\\
Research Institute for Solid State Physics and Optics\\ Hungarian Academy of
Sciences \\ P. O. Box 49, H-1525 Budapest\\ Hungary\\
E-mail: suto@szfki.hu}
\date{}
\maketitle
\begin{abstract}
\noindent
For a noninteracting Bose gas with a fixed one-body
Hamiltonian $H^0$ independent of the number of particles
we derive the inequalities $\langle N_i\rangle_N<\langle N_i\rangle_{N+1}$,
$\langle N_iN_j\rangle_N<\langle N_i\rangle_N\langle N_j\rangle_N$ for
$i\neq j$, $\partial\langle N_0\rangle_N/\partial\beta>0$ and
$\langle N_i\rangle_N^+<\langle N_i\rangle_N$.
Here $N_i$ is the occupation number of the $i\,$th eigenstate of $H^0$, $\beta$ is the
inverse temperature and the superscript + refers to adding an extra level to those of $H^0$.
The results follow from the convexity of the $N$-particle free energy as a function of $N$.
\vspace{5mm}

\\
\end{abstract}

Correlation inequalities have been playing a particular role in statistical physics. They have always
been limited to special models, most of them being valid for classical spins on lattices
\cite{Gr,Br}.
Nevertheless, when they could be applied, they proved to be very useful in deriving rigorous
results, for example, on phase transitions or on the existence of limits of correlation functions.

In this paper I present correlation inequalities for a family of continuous or lattice quantum systems:
noninteracting Bose gases in boxes, on tori or in some sufficiently fast increasing confining
external potential.
Such a system is characterized by a one-body Hamiltonian $H^0$ having a discrete spectrum
$\veps_0<\veps_1\leq\veps_2\leq\ldots$. In the continuum the number of eigenstates is infinite and we
have to impose the condition that the sum of the Boltzmann factors $x_i=\exp(-\beta\veps_i)$ is finite,
\be
\tr e^{-\beta H^0}=\sum_{i=0}^\infty e^{-\beta\veps_i}\equiv\sum_{i=0}^\infty x_i<\infty\ .
\ee
The relevant random variables are the occupation numbers $N_i$ of the eigenstates of $H^0$. In the
canonical ensemble they are weakly dependent, their only inter-dependence coming from the constraint
$\sum N_i=N$. Hence, their joint probability distribution is
\be
P_N(N_0=n_0,N_1=n_1,\ldots)=Z_N^{-1}\delta_{N,\sum n_i}\prod_{j\geq 0}x_j^{n_j}\ ,\quad
Z_N=\sum_{\{n_j\}:\sum n_j=N}\prod_{j\geq 0}x_j^{n_j}
\ee
Note that $Z_0=1$.
Let $H^0_+$ be a one-body Hamiltonian whose spectrum $\mbox{spec\/}H^0_+=\mbox{spec\/}H^0\cup\{\veps\}$.
Below
$\langle \cdot\rangle_N$ and $\langle \cdot\rangle_N^+$ denote mean values in the $N$-particle
canonical ensembles
generated by $H^0$ and $H^0_+$, respectively.

\begin{theorem}
(i) $\langle N_i\rangle_N<\langle N_i\rangle_{N+1}$\\
(ii) $\partial\langle N_i\rangle_N/\partial\ln x_j=\langle N_iN_j\rangle_N
-\langle N_i\rangle_N\langle N_j\rangle_N<0$ for $i\neq j$\\
(iii) $\partial\langle N_0\rangle_N/\partial\beta>0$\\
(iv) $\langle N_i\rangle_N^+<\langle N_i\rangle_N$.

\end{theorem}

The content of these inequalities is intuitively obvious but not more obvious than that of the
inequalities of Griffiths, Kelly and Sherman (GKS) which say that in ferromagnetic Ising models the spins
are positively correlated \cite{G2,KS}. Having in mind the subtleties of the proof of the GKS
inequalities, one cannot expect a genuinly simple proof for this theorem either.

Inequality (iv) can be put in a more general form. Let $H^1$ be a one-particle Hamiltonian,
suppose that $\exp(-\beta H^1)$ is trace class and
$\mbox{spec\/}H^0\subset\mbox{spec\/}H^1$ where repeated eigenvalues are considered separately.
If $\veps_i$ is a common eigenvalue then the
occupation number $N_i$ of the corresponding eigenstate (which may be different for $H^0$ and $H^1$)
satisfies\\

\noindent
{\em (iv')} $\langle N_i\rangle_N^1<\langle N_i\rangle_N$\\

\noindent
where $\langle\cdot\rangle_N^1$ denotes the canonical expectation value with respect to $H^1$.
One obtains (iv') by a repeated
application of (iv) or directly by a simple modification of the proof of (iv).
One may wonder about the relevance of the fourth inequality.
One-particle Hamiltonians with a modified
spectrum can serve as auxiliary tools, as for instance in \cite{S1,S2}.
Actually, the motivation of the present paper was the need of this inequality
in proving the occurrence of a generalized Bose-Einstein condensation in an interacting trapped Bose
gas, when both the interaction and the one-dimensional harmonic trap potential are scaled as $N$
tends to infinity \cite{S2}.

The basic ingredient of the proof of the theorem is the convexity of the $N$-particle free energy.

\begin{lemma}
The $N$-particle free energy is a convex function of $N$, namely
\be\label{lemma}
Z_mZ_{n+1}\leq Z_{m+1}Z_n\quad\mbox{for any}\quad m<n.
\ee
\end{lemma}
Introducing $F_N=-\ln Z_N$, (\ref{lemma}) is equivalent to $F_{m+1}-F_m\leq F_{n+1}-F_n$ which is just
convexity.
The usual statement about the convexity of the free energy in homogenous systems is that
in the thermodynamic limit
the free energy density is a convex function of the particle number density. This holds quite generally
true,
whenever there is asymptotic equivalence between canonical and grand-canonical ensembles. Indeed, in this
case the free energy density is the Legendre transform of the pressure which is trivially convex as a
function of the
chemical potential. Inequality (\ref{lemma}) is more model dependent. It perhaps remains
true in the so-called diagonal model of a Bose gas \cite{DLP} if the pair interaction has a nonnegative
Fourier transform, because then
there is a repulsion between the occupation numbers of different plane-wave states.
Also, in classical lattice models with repulsive interactions (\ref{lemma}) may hold true because dividing
a finite piece of a lattice into two parts, the more equal the sizes of the parts the larger
the number of repulsive links to be cut (and the product of the partition functions of the two parts
with it). Note that the subadditive property $\ln Z_{m+n}\leq\ln Z_n+\ln Z_m$ is a special case of
(\ref{lemma}) ($Z_0=1$).

\vspace{2mm}
\noindent
{\it Proof of the theorem.}\\
We prove the first inequality by using the lemma. The scheme of the proof of the rest will then be
$(i)\Rightarrow (ii)\Rightarrow (iii)$ and $(i)\Rightarrow (iv)$.

(i) Let us introduce the notation $Z_{N,i}$ for the $N$-particle partition function in a system where the
level $i$ is missing,
\be
Z_{N,i}=\sum_{\{n_j\}_{j\neq i}:\sum n_j=N}\prod_{j\geq 0}x_j^{n_j}
\ee
The lemma applies to these partition functions as well. For any $i$
\be\label{ZNdecomp}
Z_N=\sum_{k=0}^Nx_i^kZ_{N-k,i}
\ee
Using (\ref{ZNdecomp}) and the corresponding identity for $Z_{N+1}$,
\bea
\langle N_i\rangle_{N+1}-\langle N_i\rangle_{N}=\frac{1}{Z_{N+1}}\sum_{m=1}^{N+1}mx_i^mZ_{N+1-m,i}
-\frac{1}{Z_N}\sum_{m=1}^{N}mx_i^mZ_{N-m,i}\phantom{aaaaaaaaaaa}\nonumber\\
=\frac{1}{Z_NZ_{N+1}}\left\{\sum_{k=0}^N\sum_{m=1}^{N+1}mx_i^{k+m}Z_{N-k,i}Z_{N+1-m,i}
-\sum_{k=0}^{N+1}\sum_{m=1}^Nmx_i^{k+m}Z_{N-m,i}Z_{N+1-k,i}\right\}\nonumber\\
=\frac{1}{Z_NZ_{N+1}}\sum_{k=0}^N\sum_{m=1}^{N}mx_i^{k+m}[Z_{N-k,i}Z_{N+1-m,i}-Z_{N-m,i}Z_{N+1-k,i}]
\nonumber\\
+\frac{1}{Z_NZ_{N+1}}\left\{\sum_{k=0}^N(N+1)x_i^{k+N+1}Z_{N-k,i}-\sum_{m=1}^Nmx_i^{m+N+1}Z_{N-m,i}\right\}
\nonumber\\
=\frac{1}{Z_NZ_{N+1}}\sum_{k=1}^N\sum_{m=1}^{N}mx_i^{k+m}[Z_{N-k,i}Z_{N+1-m,i}-Z_{N-m,i}Z_{N+1-k,i}]
\nonumber\\
+\frac{1}{Z_NZ_{N+1}}\left\{\sum_{m=1}^{N}mx_i^{m}[Z_{N,i}Z_{N+1-m,i}-Z_{N-m,i}Z_{N+1,i}]\right.\nonumber\\
\left.+(N+1)x_i^{N+1}Z_{N,i}+
\sum_{m=1}^N(N+1-m)x_i^{m+N+1}Z_{N-m,i}\right\}.\nonumber
\eea
The expression in
the last line is strictly positive and, due to the lemma, in the penultimate line each term of the
single sum is nonnegative. Let us rewrite the double sum. Since the diagonal terms vanish and the
difference in the square bracket changes sign if $k$ and $m$ are interchanged, we obtain
\[\sum_{1\leq k<m\leq N}x_i^{k+m}(m-k)[Z_{N-k,i}Z_{N+1-m,i}-Z_{N-m,i}Z_{N+1-k,i}].\]
Because the smallest among the four subscripts is $N-m$, we find again with the lemma that each term is
nonnegative. This proves inequality (i).

(ii) For $i\neq j$ let us introduce
\be
\langle N_i\rangle_{n,j}=\frac{1}{Z_{n,j}}\sum_{k=1}^nkx_i^kZ_{n-k,i,j}\ ,
\ee
the mean value of $N_i$ in the $n$-particle canonical ensemble of a system in which the level $j$ is
missing. Here $Z_{n-k,i,j}$ is the $(n-k)$-particle partition function in a system with missing levels $i,j$.
For any $N\geq n$ and $m=N-n$, $\langle N_i\rangle_{n,j}$
agrees with the conditional expectation value of $N_i$
in the $N$-particle system provided that $N_j=m$.
The inequality (i) is valid also in this case. One can write
\be\label{condexp}
\langle N_i\rangle_{N}=\sum_{m=0}^Np_m(x_j)\langle N_i\rangle_{N-m,j}
\ee
where
\be
p_m(x_j)=\frac{x_j^mZ_{N-m,j}}{Z_N}
\ee
are probabilities, $\sum_{m=0}^Np_m(x_j)=1$ (and $\sum_{m=1}^Nmp_m(x_j)=\langle N_j\rangle_N$). Now
\be
\frac{\partial p_m(x_j)}{\partial\ln x_j}=\frac{x_j^mZ_{N-m,j}}{Z_N}(m-\langle N_j\rangle_N)
=p_m(x_j)(m-\langle N_j\rangle_N)
\ee
changes sign when $m$ passes through $\langle N_j\rangle_N$.
Therefore, by applying the inequality (i) to $\langle N_i\rangle_{N-m,j}$ we find
\bea\label{corr}
\frac{\partial \langle N_i\rangle_{N}}{\partial\ln x_j}&=&
\langle N_iN_j\rangle_N-\langle N_i\rangle_N\langle N_j\rangle_N=\sum_{m=0}^N
\frac{\partial p_m(x_j)}{\partial\ln x_j}\langle N_i\rangle_{N-m,j}\nonumber\\
&=&\sum_{0\leq m<\langle N_j\rangle_N}\frac{\partial p_m(x_j)}{\partial\ln x_j}\langle N_i\rangle_{N-m,j}
+\sum_{\langle N_j\rangle_N<m\leq N}\frac{\partial p_m(x_j)}{\partial\ln x_j}\langle N_i\rangle_{N-m,j}
\nonumber\\
&<&\langle N_i\rangle_{N-\lfloor\langle N_j\rangle_N\rfloor,j}\sum_{0\leq m<\langle N_j\rangle_N}
\frac{\partial p_m(x_j)}{\partial\ln x_j}+\langle N_i\rangle_{N-\lceil\langle N_j\rangle_N\rceil,j}
\sum_{\langle N_j\rangle_N<m\leq N}\frac{\partial p_m(x_j)}{\partial\ln x_j}\nonumber\\
&\leq&\langle N_i\rangle_{N-\lfloor\langle N_j\rangle_N\rfloor,j}\frac{\partial}{\partial\ln x_j}
\sum_{m=0}^Np_m(x_j)=0
\eea
which proves the second inequality. The identity
\be
\langle N_iN_j\rangle_N=\sum_{m=1}^Nmp_m(x_j)\langle N_i\rangle_{N-m,j}
\ee
can be read off from the first line of (\ref{corr}).

(iii)
\bea
\frac{\partial\langle N_0\rangle_N}{\partial\beta}=-\left\langle N_0\sum N_j\veps_j\right\rangle_N
+\langle N_0\rangle_N\left\langle\sum N_j\veps_j\right\rangle_N
=-\sum_{j=0}^\infty\veps_j\left[\langle N_0 N_j\rangle_N-\langle N_0\rangle_N\langle N_j\rangle_N\right]
\nonumber\\
=-\veps_0[\langle N_0^2\rangle_N-\langle N_0\rangle_N^2)]
-\sum_{j=1}^\infty\veps_j[\langle N_0 N_j\rangle_N-\langle N_0\rangle_N\langle N_j\rangle_N]\nonumber\\
=-\sum_{j=1}^\infty(\veps_j-\veps_0)[\langle N_0 N_j\rangle_N-\langle N_0\rangle_N\langle N_j\rangle_N]
>0
\eea
where we used
\be
\sum_{j=0}^\infty[\langle N_0 N_j\rangle_N-\langle N_0\rangle_N\langle N_j\rangle_N]
=N[\langle N_0\rangle_N-\langle N_0\rangle_N]=0
\ee
in the last equality and (ii) in the inequality.

(iv) Let $x=\exp(-\beta\veps)$ and let $Z_N^+$ denote the $N$-particle canonical partition function
of the system with the additional level. In analogy with (\ref{condexp}) and by applying (i)
\be
\langle N_i\rangle_N^+=\sum_{m=0}^N\frac{x^mZ_{N-m}}{Z_N^+}\langle N_i\rangle_{N-m}
\equiv\sum_{m=0}^Np_m^+(x)\langle N_i\rangle_{N-m}<\langle N_i\rangle_{N}
\ee
because $p_0^+(x)=Z_N/Z_N^+<1$.

This finishes the proof of the theorem.

\vspace{2mm}
\noindent
{\it Proof of the lemma.}\\
Let $p_1\leq p_2\leq\cdots\leq p_l$ be a partition of $N$, that is, $p_1\geq 1$ and $\sum_1^l p_i=N$.
We denote this by
$\pv=(p_1,p_2,\ldots,p_l)\vdash N$. For each $m\in [0,N]$, $Z_mZ_{N-m}$ is a sum over the same set of
monomials of $x_1,x_2,\ldots$, namely all those $\prod_j x_j^{n_j}$ with $\sum_jn_j=N$. For a given
$m$ the coefficient of a monomial depends only on the set of nonzero $n_j$ and not on their order or
on the set of the corresponding subscripts $j$. Therefore
\be
Z_mZ_{N-m}=\sum_{\pv\vdash N}a(\pv|m)\sum_{\{n_j\}\sim\pv}\prod x_j^{n_j}
\ee
where $\{n_j\}\sim\pv$ means that, after rearranging in increasing order, the vector of nonzero $n_j$
agrees with $\pv$.
We shall prove that for any $1\leq p_1\leq\cdots\leq p_l$
\be\label{ineq}
a(p_1,p_2,\ldots,p_l|m)\leq a(p_1,p_2,\ldots,p_l|m+1)\ \ {\rm if}\ \  m<\frac{1}{2}(p_1+\cdots+p_l)
\ee
which implies (\ref{lemma}) through a term-by-term inequality. The proof is obtained by induction over $l$.

$a(p_1,\ldots,p_l|m)$
is the number of different $m$th degree monomials within $x_1^{p_1}\cdots x_l^{p_l}$ or, in
more popular terms, the number of possibilities to distribute $m$ Euros among $l$ people in such a way
that they can obtain at most $p_1,\ldots,p_l$, respectively. Thus, $a(p|m)=1$ for any $p\geq 1$ and any
$0\leq m\leq p$, so that for $l=1$ (\ref{ineq}) holds with equality. The induction hypothesis is that
for an $l>1$
\be\label{indhyp}
a(p_1,\ldots,p_{l-1}|m)\leq a(p_1,\ldots,p_{l-1}|m+1)\ \ {\rm if}\ \  m<\frac{1}{2}(p_1+\cdots+p_{l-1}).
\ee
Because of the identity
\be
a(\pv|m)=a(\pv|\sum p_i-m)
\ee
the hypothesis (\ref{indhyp}) is equivalent to
\be
a(p_1,\ldots,p_{l-1}|m)\leq a(p_1,\ldots,p_{l-1}|n)\ \ {\rm if}\ \  m\leq\min\{n,\ p_1+\cdots+p_{l-1}-n\}.
\ee
Let us introduce the notations
\be
Q(p_1,\ldots,p_l)=\{\q\in\Z^l:0\leq q_i\leq p_i,\  1\leq i\leq l\}\ ,
\ee
\be
C(p_1,\ldots,p_l|m)=\{\q\in Q(p_1,\ldots,p_l):\sum_1^lq_i=m\}
\ee
and
\be\label{pcm}
PC(p_1,\ldots,p_l|m)=\{\q\in Q(p_1,\ldots,p_{l-1}):m-p_l\leq\sum_1^{l-1}q_i\leq m\}.
\ee
We have
\be
a(p_1,\ldots,p_l|m)=|C(p_1,\ldots,p_l|m)|=|PC(p_1,\ldots,p_l|m)|
\ee
$|A|$ meaning the number of elements of the set $A$.
The first equality is obvious. The second comes from the fact that $PC(\pv|m)$ is the orthogonal
projection of $C(\pv|m)$ onto the $l-1$ dimensional subspace perpendicular to the $l$\,th unit vector and,
because of the constraint $\sum_1^l q_i=m$, the projection is a bijection.
Comparing (\ref{pcm}) with
\be\label{pcm+1}
PC(p_1,\ldots,p_l|m+1)=\{\q\in Q(p_1,\ldots,p_{l-1}):m+1-p_l\leq\sum_1^{l-1}q_i\leq m+1\}
\ee
we find that
\bea\label{1}
|PC(\pv|m+1)\setminus PC(\pv|m)|&=&
|\{\q\in Q(p_1,\ldots,p_{l-1}): \sum_1^{l-1}q_i= m+1\}|\nonumber\\
&=&a(p_1,\ldots,p_{l-1}|m+1)
\eea
while
\bea\label{2}
|PC(\pv|m)\setminus PC(\pv|m+1)|&=&
|\{\q\in Q(p_1,\ldots,p_{l-1}): \sum_1^{l-1}q_i= m-p_l\}|\nonumber\\
&=&a(p_1,\ldots,p_{l-1}|m-p_l).
\eea
Note that either of (\ref{1}) and (\ref{2}) can vanish if $m+1>p_1+\cdots+p_{l-1}$ or $m-p_l<0$,
respectively. The inequality (\ref{ineq}) is satisfied if
\be
a(p_1,\ldots,p_{l-1}|m-p_l)\leq a(p_1,\ldots,p_{l-1}|m+1)\ ,
\ee
and this follows from the induction hypothesis if
\be\label{min}
m-p_l\leq\min \{m+1,\ p_1+\cdots+p_{l-1}-(m+1)\}.
\ee
If $m+1\leq (p_1+\cdots+p_{l-1})/2$ then this is the minimum on the right-hand side of (\ref{min}) which
is therefore fulfilled. If $m+1>(p_1+\cdots+p_{l-1})/2$,
the minimum is $p_1+\cdots+p_{l-1}-(m+1)$. But then
\[m-p_l\leq p_1+\cdots+p_{l-1}-(m+1)\]
is equivalent to
\[m+\frac{1}{2}\leq \frac{1}{2}(p_1+\cdots+p_l)\]
which holds true because $m<(p_1+\cdots+p_l)/2$. By this we have finished the proof.

\vspace{2mm}
Concerning the proof of the lemma, two remarks are in order. First, it is tempting to present a more
elegant argument which refers to the convexity of the hyper-rectangle $Q(\pv)$. If we had $\R^l$ instead
of $\Z^l$ in the definition of $Q(\pv)$, $C(\pv|m)$ would be the intersection of the rectangle with the
hyperplane $\sum q_i=m$. Because of the convexity of $Q(\pv)$ the $l-1$ dimensional Lebesgue measure of
$C(\pv|m)$ cannot have a minimum in the interval $0<m<\sum p_i$. By symmetry,
it has a not necessarily strict
maximum at $m=\sum p_i/2$, implying that it is nondecreasing for
$m<\sum p_i/2$. However, we do
not need this result for the Lebesgue measure of the cut but for the number of points of integer
coordinates on it, which makes the convexity argument somewhat shaky.

Second, the number in question can be given by a formula. It reads
\be
a(\pv|m)=\sum_{q_1=0}^{p_1}\cdots\sum_{q_j=\max\{0,m-\sum_{i=1}^{j-1}q_i-\sum_{i=j+1}^lp_i\}}
^{\min\{p_j,m-\sum_{i=1}^{j-1}q_i\}}\cdots
\sum_{q_{l-1}=\max\{0,m-\sum_{i=1}^{l-2}q_i-p_l\}}
^{\min\{p_{l-1},m-\sum_{i=1}^{l-2}q_i\}}1\ .
\ee
If $m\leq p_l$, in all the summations the lower bounds equal zero and thus for $m<p_l$,
$a(\pv|m)\leq a(\pv|m+1)$ can explicitly be seen. For $m\leq p_i$ by simple combinatorial
considerations we find
\be
a(\pv|m)=\sum_{q_1=0}^{p_1}\cdots\sum_{q_{i-1}=0}
^{\min\{p_{i-1},m-\sum_{j=1}^{i-2}q_j\}}{m-\sum_{j=1}^{i-1}q_{j}+l-i\choose l-i}\ .
\ee
If $m>p_l$ the general formula is less transparent. This is why we have opted
for the inductive proof.

\vspace{2mm}
\noindent
{\bf Acknowledgment} This work has benefited from the support of the Hungarian Scientific Research Fund through
Grant T 42914.

\vspace{2mm}
\noindent
{\bf Note added.} After having submitted the paper, I learned from Valentin Zagrebnov that
the convexity of the free energy was earlier shown by Lewis, Zagrebnov and Pul\'e \cite{LZP}. Their proof
is somewhat shorter, the present one provides somewhat more information due to the established connection
with combinatorics and geometry. Also, I was informed about a recent work by Pul\'e and Zagrebnov
\cite{PZ} in which inequality (i) is derived and used.

\end{document}